

\def\singlespace{\normalbaselines}
\def\oneandahalfspace{\baselineskip=1.15\normalbaselineskip plus 1pt
\lineskip=2pt\lineskiplimit=1pt}

\def\np{\vfill\eject}
\def\nl{\hfil\break}

\def\nofirstpagenoten{\nopagenumbers\footline={\ifnum\pageno>1\tenrm
\hss\folio\hss\fi}}
\def\nofirstpagenotwelve{\nopagenumbers\footline={\ifnum\pageno>1\twelverm
\hss\folio\hss\fi}}
\def\leaderfill{\leaders\hbox to 1em{\hss.\hss}\hfill}
\def\ft#1#2{{\textstyle{{#1}\over{#2}}}}
\def\frac#1/#2{\leavevmode\kern.1em
\raise.5ex\hbox{\the\scriptfont0 #1}\kern-.1em/\kern-.15em
\lower.25ex\hbox{\the\scriptfont0 #2}}
\def\sfrac#1/#2{\leavevmode\kern.1em
\raise.5ex\hbox{\the\scriptscriptfont0 #1}\kern-.1em/\kern-.15em
\lower.25ex\hbox{\the\scriptscriptfont0 #2}}


\parindent=20pt
\def\narrow{\advance\leftskip by 40pt \advance\rightskip by 40pt}

\def\AB{\bigskip
        \centerline{\bf ABSTRACT}\medskip\narrow}
\def\nonarrower{\advance\leftskip by -40pt\advance\rightskip by -40pt}
\def\AE{\bigskip\nonarrower}

\def\boxit#1{\vbox{\hrule\hbox{\vrule\kern3pt
        \vbox{\kern3pt#1\kern3pt}\kern3pt\vrule}\hrule}}

\def\gtorder{\mathrel{\raise.3ex\hbox{$>$}\mkern-14mu
             \lower0.6ex\hbox{$\sim$}}}
\def\ltorder{\mathrel{\raise.3ex\hbox{$<$}|mkern-14mu
             \lower0.6ex\hbox{\sim$}}}
\def\dalemb#1#2{{\vbox{\hrule height .#2pt
        \hbox{\vrule width.#2pt height#1pt \kern#1pt
                \vrule width.#2pt}
        \hrule height.#2pt}}}

\font\fourteentt=cmtt10 scaled \magstep2
\font\fourteenbf=cmbx12 scaled \magstep1
\font\fourteenrm=cmr12 scaled \magstep1
\font\fourteeni=cmmi12 scaled \magstep1
\font\fourteenss=cmss12 scaled \magstep1
\font\fourteensy=cmsy10 scaled \magstep2
\font\fourteensl=cmsl12 scaled \magstep1
\font\fourteenex=cmex10 scaled \magstep2
\font\fourteenit=cmti12 scaled \magstep1
\font\twelvett=cmtt10 scaled \magstep1 \font\twelvebf=cmbx12
\font\twelverm=cmr12 \font\twelvei=cmmi12
\font\twelvess=cmss12 \font\twelvesy=cmsy10 scaled \magstep1
\font\twelvesl=cmsl12 \font\twelveex=cmex10 scaled \magstep1
\font\twelveit=cmti12
\font\tenss=cmss10
 
 \font\ninebf=cmbx7 scaled \magstep1
\font\ninerm=cmr7 scaled \magstep1 \font\ninei=cmmi7 scaled \magstep1
\font\ninesy=cmsy7 scaled \magstep1 
\font\eightrm=cmr7 scaled 1140 
 
\font\sevenbf=cmbx7 \font\sevenrm=cmr7 \font\seveni=cmmi7
\font\sevensy=cmsy7 

\catcode`@=11
\newskip\ttglue
\newfam\ssfam

\def\fourteenpoint{\def\rm{\fam0\fourteenrm}
\textfont0=\fourteenrm \scriptfont0=\tenrm \scriptscriptfont0=\sevenrm
\textfont1=\fourteeni \scriptfont1=\teni \scriptscriptfont1=\seveni
\textfont2=\fourteensy \scriptfont2=\tensy \scriptscriptfont2=\sevensy
\textfont3=\fourteenex \scriptfont3=\fourteenex \scriptscriptfont3=\fourteenex
\def\it{\fam\itfam\fourteenit} \textfont\itfam=\fourteenit
\def\sl{\fam\slfam\fourteensl} \textfont\slfam=\fourteensl
\def\bf{\fam\bffam\fourteenbf} \textfont\bffam=\fourteenbf
\scriptfont\bffam=\tenbf \scriptscriptfont\bffam=\sevenbf
\def\tt{\fam\ttfam\fourteentt} \textfont\ttfam=\fourteentt
\def\ss{\fam\ssfam\fourteenss} \textfont\ssfam=\fourteenss
\tt \ttglue=.5em plus .25em minus .15em
\normalbaselineskip=16pt
\abovedisplayskip=16pt plus 4pt minus 12pt
\belowdisplayskip=16pt plus 4pt minus 12pt
\abovedisplayshortskip=0pt plus 4pt
\belowdisplayshortskip=9pt plus 4pt minus 6pt
\parskip=5pt plus 1.5pt
\setbox\strutbox=\hbox{\vrule height12pt depth5pt width0pt}
\let\sc=\tenrm
\let\big=\fourteenbig \normalbaselines\rm}
\def\fourteenbig#1{{\hbox{$\left#1\vbox to12pt{}\right.\n@space$}}}

\def\twelvepoint{\def\rm{\fam0\twelverm}
\textfont0=\twelverm \scriptfont0=\ninerm \scriptscriptfont0=\sevenrm
\textfont1=\twelvei \scriptfont1=\ninei \scriptscriptfont1=\seveni
\textfont2=\twelvesy \scriptfont2=\ninesy \scriptscriptfont2=\sevensy
\textfont3=\twelveex \scriptfont3=\twelveex \scriptscriptfont3=\twelveex
\def\it{\fam\itfam\twelveit} \textfont\itfam=\twelveit
\def\sl{\fam\slfam\twelvesl} \textfont\slfam=\twelvesl
\def\bf{\fam\bffam\twelvebf} \textfont\bffam=\twelvebf
\scriptfont\bffam=\ninebf \scriptscriptfont\bffam=\sevenbf
\def\tt{\fam\ttfam\twelvett} \textfont\ttfam=\twelvett
\def\ss{\fam\ssfam\twelvess} \textfont\ssfam=\twelvess
\tt \ttglue=.5em plus .25em minus .15em
\normalbaselineskip=14pt
\abovedisplayskip=14pt plus 3pt minus 10pt
\belowdisplayskip=14pt plus 3pt minus 10pt
\abovedisplayshortskip=0pt plus 3pt
\belowdisplayshortskip=8pt plus 3pt minus 5pt
\parskip=3pt plus 1.5pt
\setbox\strutbox=\hbox{\vrule height10pt depth4pt width0pt}
\let\sc=\ninerm
\let\big=\twelvebig \normalbaselines\rm}
\def\twelvebig#1{{\hbox{$\left#1\vbox to10pt{}\right.\n@space$}}}

\def\tenpoint{\def\rm{\fam0\tenrm}
\textfont0=\tenrm \scriptfont0=\sevenrm \scriptscriptfont0=\fiverm
\textfont1=\teni \scriptfont1=\seveni \scriptscriptfont1=\fivei
\textfont2=\tensy \scriptfont2=\sevensy \scriptscriptfont2=\fivesy
\textfont3=\tenex \scriptfont3=\tenex \scriptscriptfont3=\tenex
\def\it{\fam\itfam\tenit} \textfont\itfam=\tenit
\def\sl{\fam\slfam\tensl} \textfont\slfam=\tensl
\def\bf{\fam\bffam\tenbf} \textfont\bffam=\tenbf
\scriptfont\bffam=\sevenbf \scriptscriptfont\bffam=\fivebf
\def\tt{\fam\ttfam\tentt} \textfont\ttfam=\tentt
\def\ss{\fam\ssfam\tenss} \textfont\ssfam=\tenss
\tt \ttglue=.5em plus .25em minus .15em
\normalbaselineskip=12pt
\abovedisplayskip=12pt plus 3pt minus 9pt
\belowdisplayskip=12pt plus 3pt minus 9pt
\abovedisplayshortskip=0pt plus 3pt
\belowdisplayshortskip=7pt plus 3pt minus 4pt
\parskip=0.0pt plus 1.0pt
\setbox\strutbox=\hbox{\vrule height8.5pt depth3.5pt width0pt}
\let\sc=\eightrm
\let\big=\tenbig \normalbaselines\rm}
\def\tenbig#1{{\hbox{$\left#1\vbox to8.5pt{}\right.\n@space$}}}
\let\rawfootnote=\footnote \def\footnote#1#2{{\rm\parskip=0pt\rawfootnote{#1}
{#2\hfill\vrule height 0pt depth 6pt width 0pt}}}

\def\tenfoot{\tenpoint\hskip-\parindent\hskip-.1cm}

\overfullrule=0pt
\twelvepoint
\def\sbullet{\raise.2em\hbox{$\scriptscriptstyle\bullet$}}
\nofirstpagenotwelve
\hsize=16.5 truecm
\baselineskip 15pt

\def\ft#1#2{{\textstyle{{#1}\over{#2}}}}

\def\a{\alpha_0}
\def\lra{\leftrightarrow}
\def\b{\beta}
\def\bb{\bar\beta}

\def\t{\theta}

\def\del{\partial}

\def\tW{\widetilde W}
\def\tU{\widetilde U}
\def\tV{\widetilde V}

\def\cramp{\medmuskip = -2mu plus 1mu minus 2mu}
\def\uncramp{\medmuskip = 4mu plus 2mu minus 4mu}

\oneandahalfspace
\rightline{CTP TAMU--85/91}
\rightline{October 1991}

\vskip 2truecm
\centerline{\bf Anomaly Freedom and Realisations for Super-$W_3$ Strings}
\vskip 1.5truecm
\centerline{H. Lu,$^\star$ C.N. Pope,\footnote{$^
\star$}{\tenfoot Supported in part by the
U.S. Department of Energy, under
grant DE-FG05-91ER40633.} X.J. Wang and K.W.
Xu.\footnote{$^\$$}{\tenfoot Supported by a World Laboratory
Scholarship.}}
\vskip 1.5truecm
\centerline{\it Center
for Theoretical Physics,
Texas A\&M University,}
\centerline{\it College Station, TX 77843--4242, USA.}

\vskip 1.5truecm
\AB\singlespace

      We construct new multi-field realisations of the $N=2$ super-$W_3$
algebra, which are important for building super-$W_3$ string theories.
We derive the structure of the ghost vacuum for such theories, and use the
result to calculate the intercepts. These results determine the conditions
for physical states in the super-$W_3$ string theory.

\AE\oneandahalfspace

\np
\noindent
{\bf 1. Introduction}
\bigskip

     Two-dimensional gravity is the gauge theory of the Virasoro algebra. In
general, when this theory is quantised, the gauge symmetry is anomalous. The
condition for anomaly freedom is that the matter fields should realise the
Virasoro algebra with central charge $c=26$. A convenient way to describe
this is in the BRST formalism; the condition for anomaly freedom is
equivalent to the requirement that the BRST charge $Q$ be nilpotent. In fact
the nilpotency condition determines not only that $c=26$ but also that the
intercept for $L_0$ is $1$. In a critical string theory, the $c=26$
matter system is realised by scalar fields which are interpreted as
space-time coordinates. The physical states of the theory must satisfy the
conditions
$$\eqalign{
(L_0-1)\big|{\rm phys}\big\rangle&=0\ ,\cr
L_n\big|{\rm phys}\big\rangle&=0\ ,\qquad n>0\ .\cr}\eqno(1.1)
$$

     There exist various higher-spin extensions of the Virasoro algebra,
known as $W$ algebras. The gauge theories of these algebras give rise to
extensions of two-dimensional gravity, known as $W$ gravities. The quantum
$W$ gravity theories can again be conveniently described in the BRST
formalism. Once more, anomaly freedom is guaranteed by the nilpotency of the
BRST charge $Q$, which determines the central charge of the matter system.
The nilpotency also determines the intercepts for the zero Laurent modes of
all the currents in the algebra. For example, in the $W_3$ case [1], which has
currents $T$ and $W$ of spins 2 and 3, the nilpotency of $Q$ implies that
the central charge $c=100$, and that the intercepts for $L_0$ and $W_0$ are
4 and 0 respectively [2]. A two-scalar realisation of $W_3$ at arbitrary
$c$, coming from the quantum Miura transformation, is given in [3]. Recently,
a generalisation to an $n$-scalar realisation has been given in [4]. Using
these results, anomaly-free $W_3$ gravity and critical $W_3$ string theory
have been constructed [5,6]. It appears that the physical spectrum of the
$W_3$ string contains no massless states [7].

    It is natural to consider supersymmetric extensions of the $W$ gravities
and strings. Recently, the $N=2$ super-$W_3$ algebra has been constructed;
classically in [8], and at the quantum level in [9]. The quantum algebra exists
for arbitrary values of the central charge $c$. A realisation of this algebra
in terms of two complex $N=2$ superfields has recently been constructed, for
arbitrary $c$, by using the super-Miura transformation [10].  The spin
content of the $N=2$ super-$W_3$ algebra, grouped into $N=2$
supermultiplets, is
$$
\left\{\matrix{&\ft32&\cr
               1&&2\cr
               &\ft32&\cr}\right\}
\left\{\matrix{&\ft52&\cr
               2&&3\cr
               &\ft52&\cr}\right\}.\eqno(1.2)
$$
Our notation for the currents of the $N=2$ super-$W_3$ algebra is:
$$
\left\{\matrix{&G^+&\cr
               J&&T\cr
               &G^-&\cr}\right\}
\left\{\matrix{&U^+&\cr
               V&&W\cr
               &U^-&\cr}\right\}.\eqno(1.3)
$$
The generalisation to the $N=2$ super-$W_n$ algebra has the spin content
$$
\left\{\matrix{&\ft32&\cr
               1&&2\cr
               &\ft32&\cr}\right\}
\left\{\matrix{&\ft52&\cr
               2&&3\cr
               &\ft52&\cr}\right\}
\left\{\matrix{&\ft72&\cr
               3&&4\cr
               &\ft72&\cr}\right\}
\cdots
\left\{\matrix{&(n-\ft12)&\cr
               (n-1)&&n\cr
               &(n-\ft12)&\cr}\right\}.\eqno(1.4)
$$

     In this paper we begin by reviewing the derivation of the
two-complex-superfield realisation from the super-Miura transformation.
Using this we then construct new realisations of $N=2$ super-$W_3$ in terms
of arbitrary numbers of superfields, for general values of $c$. These
realisations are important for the construction of super-$W_3$ strings. In
order to construct such theories that are anomaly free, it would be useful to
know the BRST operator $Q$ for the algebra. In particular, one could deduce
the critical central charge $c$ and the intercepts for the matter currents
from the nilpotency of $Q$. The form of the BRST operator for this algebra
is presently unknown; even for the relatively simple $W_3$ algebra the BRST
operator is rather complicated [2]. Fortunately there are ways to determine
the central charge and the intercepts without needing to know the explicit
expression for the BRST operator. The critical central charge has been shown
to be $c=12$ [9]. In this paper, we show that the intercepts for the
bosonic currents are all zero.

     The organisation of this paper is as follows. In section 2, we review
the construction of the two-complex superfield realisation of the super-$W_3$
algebra from the quantum Miura transformation. In section 3, we generalise
this to multi-field realisations. In section 4 we derive the structure of the
ghost vacuum for the $N=2$ super-$W_n$ algebras, and hence show that the
$L_0$ intercept vanishes for all $n$.  In section 5, we calculate the
intercepts for the remaining currents of the super-$W_3$ algebra, by
examining certain null states. The paper ends with conclusions and
discussions in section 6.

\bigskip

\noindent
{\bf 2. Miura Realisation for the Super-$W_3$ Algebra}
\bigskip

      In this section, we review the discussion of ref.\ [10], where the
two-complex-superfield realisation of $N=2$ super-$W_3$ algebra was derived.
In fact for the first part of this discussion we may consider the more
general case of the $N=2$ super-$W_n$ algebra. Although we shall
subsequently adopt an $N=2$ superfield notation, the Miura transformation
is most conveniently expressed in terms of $N=1$ superfields [10].

      Let $L$ be the differential operator
$$
L=u_n D^n+u_{n-2}(z,\theta) D^{n-2}+\cdots+u_1(z,\theta) D
+u_0(z,\theta),\eqno(2.1)
$$
where $z$ and $\theta$ are the bosonic and fermionic coordinates of a
2-dimensional superspace, and $D$ is given by
$$
D={\partial\over{\partial\theta}}+\theta\partial,\eqno(2.2)
$$
with $\partial\equiv\partial/\partial z$.  The derivative $D$ satisfies
$D^2=\partial$.

The Miura transformation for the super Lie algebra $A(n,n-1)$ has the form
[10]
$$
L=\Big(\prod_{i=1}^n [(\alpha_0 D+D \Phi_{n+1-i}-\chi_{i-1})(\alpha_0 D+D
\Phi_{n+1-i}-\chi_i)]\Big)(\alpha_0 D-\chi_n)\ ,\eqno(2.3)
$$
where
$$\eqalign{
\chi_0&=0\cr
\chi_i&=\sum_{k=1}^i D {\bar\Phi}_i(z, \theta)\ ,\cr}\eqno(2.4)
$$
and the $N=1$ superfields $\Phi_i(z,\theta)$ and ${\bar\Phi}_i(z,\theta)$ are
given in terms of components by
$$\eqalign{
\Phi_i(z,\theta)&=\phi_i(z)+\theta\psi_i(z)\ ,\cr
{\bar\Phi_i}(z,\theta)&={\bar\phi}_i(z)+\theta{\bar\psi}_i(z)\ .\cr}\eqno(2.5)
$$
The parameter $\alpha_0$ is related to the background charge of the
Feigin-Fuchs representation. Expanding out (2.3) in powers of $D$, and
comparing with (2.1), one can read off the superfields $u_i(z,\theta)$.

     Specialising to $n=2$, one finds that the $u_i$ for super-$W_3$ are given
by [10]:
$$\eqalign{
u_3(z,\theta) \ & = -D\Phi_1 D\bar \Phi_1
-D\Phi_2D\bar \Phi_2 +\alpha_0 D^2\Phi_1 -\alpha_0 D^2 \bar\Phi_1 +\alpha_0
D^2\Phi_2 -2\alpha_0 D^2\bar \Phi_2\ , \cr
 u_2(z,\theta)\ & =  -D^2\Phi_1\bar D \Phi_1 -D^2\Phi_2 D\bar\Phi_2-
\alpha_0 D^3\bar\Phi_1 -2\alpha_0 D^3 \bar\Phi_2\ , \cr
 u_1(z,\theta) \ & =  \alpha_0^3 D^4\Phi_1
-\alpha_0^3 D^4\bar\Phi_1-\alpha_0^3 D^4\bar\Phi_2
 -\alpha_0^2 D^3 \Phi_1
D\bar\Phi_1 - \alpha_0^2D\Phi_1D^3\bar\Phi_1 \cr
&\quad -\alpha_0^2
D^3\Phi_1D\bar\Phi_2
-\alpha_0^2 D^2\Phi_2 D^2\bar\Phi_1
-\alpha_0^2 D^2 \Phi_2 D^2 \bar\Phi_2 + \alpha_0 D^2 \bar\Phi_2
D^2 \bar\Phi_1  \cr
&\quad +\alpha_0^2 D^2 \bar\Phi_2 D^2 \bar\Phi_2 + \alpha_0^2
D^2
\Phi_2 D^2 \Phi_1 - \alpha_0^2  D^2 \bar\Phi_2 D^2 \Phi_1
- \alpha_0 D^2\Phi_2D\Phi_1D\bar\Phi_1 \cr
&\quad + \alpha_0 D^2\bar\Phi_2 D\Phi_1
D\bar\Phi_1
 +\alpha_0 D^2\bar\Phi_1 D\Phi_2 D\bar\Phi_2  + \alpha_0 D^2\bar
\Phi_2
D\Phi_2 D \bar\Phi_2 \cr
&\quad  + \alpha_0 D^2\bar\Phi_1 D\bar\Phi_2 D\Phi_1 -\alpha_0 D^2\Phi_1
D\Phi_2D\bar\Phi_2 + D\Phi_1 D\bar\Phi_1D\Phi_2 D\bar\Phi_2 \ ,\cr
 u_0(z,\theta)\  & =  -\alpha_0^3 D^5 \bar\Phi_1 -\alpha_0^3 D^5 \Phi_2 -
\alpha_0^2 D^4 \Phi_1 D \bar\Phi_1 - \alpha_0^2 D^4 \Phi_1 D\bar \Phi_2
-\alpha_0^2 D^2\Phi_1 D^3\bar\Phi_2 \cr
 &\quad -\alpha_0^2 D^2 \Phi_2 D^3 \bar\Phi_1
-\alpha_0^2 D^2\Phi_2 D^3 \bar\Phi_2
+\alpha_0^2 D^2\bar\Phi_2D^3\bar\Phi_1 + 2\alpha_0^2
D^3\bar\Phi_2D^2\bar\Phi_2 \cr
 &\quad -\alpha_0^2 D^3 \bar\Phi_1 D^2\Phi_1
+ \alpha_0 D\Phi_2D\bar\Phi_2 D^3\bar\Phi_2
+\alpha_0^2 D^3\bar\Phi_2 D^2 \bar\Phi_1
 + \alpha_0 D^3 \bar\Phi_2  D\Phi_1 D\bar\Phi_1 \cr
 &\quad +\alpha_0
D^3\bar\Phi_1 D\Phi_2D\bar\Phi_2
 - \alpha_0 D^3\bar\Phi_1 D\Phi_1 D\bar\Phi_2
-\alpha_0
D^2\Phi_2D^2\Phi_1 D\bar\Phi_1 \cr
&\quad-\alpha_0 D^2\Phi_2 D^2\Phi_1 D\bar\Phi_2
 +\alpha_0 D^2\bar\Phi_2 D^2\Phi_1 D\bar\Phi_1
 -\alpha_0 D^2\Phi_1 D^2\bar\Phi_1 D\bar\Phi_2 \cr
 &\quad+ \alpha_0 D^2 \Phi_2 D^2\bar\Phi_1 D \bar\Phi_2 +\alpha_0
D^2\bar\Phi_2D^2\Phi_2 D\bar\Phi_2
 + D^2\Phi_1D\bar\Phi_1D\Phi_2D\bar\Phi_2  \cr
&\quad +D^2\Phi_2 D\bar\Phi_2 D\Phi_1
D\bar\Phi_1 \ , \cr}\eqno(2.6)
$$
after rescaling $u_3$ and $u_2$ by a factor of $\a^{-3}$, and $u_1$ and
$u_0$ by $\a^{-1}$.
We now introduce a second anticommuting coordinate $\tilde\theta$, and
define $N=2$ chiral superfields
$$
\eqalign{
\Phi^+_i(z,\theta^+,\theta^-)&=(1-i \tilde\theta D) \Phi_i(z,\theta)=
\phi_i(z) +\sqrt2 \theta^- \psi_i(z) -\theta^+\theta^- \partial\phi_i(z),\cr
\Phi^-_i(z,\theta^+,\theta^-)&=(1+i \tilde\theta D)
\bar\Phi_i(z,\theta)=\bar\phi_i(z) +\sqrt2 \theta^+ \bar\psi_i(z)
+\theta^+\theta^- \partial\bar\phi_i(z),\cr}\eqno(2.7)
$$
where $i=1,2$, and
$$
\theta^\pm\equiv {1\over \sqrt2}\big(\theta\pm i\tilde\theta\big).\eqno(2.8)
$$
These superfields satisfy the chirality conditions
$$
D^+\Phi^-_i=0=D^-\Phi^+_i,\eqno(2.9)
$$
where
$$
D^\pm\equiv {1\over\sqrt2}\big( D_\theta\pm i
D_{\tilde\theta}\big)={\partial\over\partial\theta^\mp} +\theta^\pm \partial,
\eqno(2.10)
$$
with $D_\theta$ and $D_{\tilde\theta}$ given by (2.2).  The derivatives
$D^+$ and $D^-$ satisfy $\{D^+,D^-\}=2\partial$.

     The fields $u_i(z,\theta)$ ($i=0,1,2,3$) can be assembled into the
$N=2$ supercurrents $T(z,\theta^+,\theta^-)$ and $\widetilde
W(z,\theta^+,\theta^-)$ of the super-$W_3$ algebra:
$$
\eqalignno{
T(z,\theta^+,\theta^-)&=\ft14 u_3(z,\theta) +\ft14 i \tilde\theta
\big(2u_2(z,\theta) -D u_3(z,\theta)\big)\cr\cr
&=-\ft14 D^+ \Phi^+_1 D^- \Phi^-_1 -\ft14 D^+ \Phi^+_2 D^-\Phi^-_2\cr
&\quad +{\alpha_0\over2} \partial\Phi^+_1-{\alpha_0\over2} \partial\Phi^-_1
+{\alpha_0\over2} \partial\Phi^+_2-\alpha_0 \partial\Phi^-_2\ ,&(2.11a)\cr
\cr
{\tW}(z,\theta^+,\theta^-)&=\ft14 u_1(z,\theta) +\ft14 i
\tilde\theta \big(2 u_0(z,\theta)-D u_1(z,\theta)\big)\cr\cr
&=  {\alpha_0^3 \over 4}\del^2\Phi_1^+
-{ \alpha_0^3 \over 4} \del^2\Phi_1^--{ \alpha_0^3 \over 4 } \del^2\Phi_2^-
 -{\alpha_0^2 \over 8} D^+\del \Phi_1^+
D^-\Phi_1^- \cr
&\quad - { \alpha_0^2\over 8 }D^+\Phi_1^+\del D^-\Phi_1^-  -{\alpha_0^2 \over 8
} D^+\del\Phi_1^+D^-\Phi_2^-
-{ \alpha_0^2\over 4 }  \del\Phi_2^+ \del\Phi_1^-\cr
&\quad -{ \alpha_0^2\over 4 } \del \Phi_2^+ \del\Phi_2^- +
{\alpha_0^2  \over 4 }
\del\Phi_2^- \del\Phi_1^-
+{ \alpha_0^2 \over 4 }\del\Phi_2^-\del\Phi_2^-\cr
&\quad + { \alpha_0^2 \over 4
} \del\Phi_2^+\del\Phi_1^+ - { \alpha_0^2 \over 4 } \del\Phi_2^-
\del\Phi_1^+
- { \alpha_0 \over 8 }\del\Phi_2^+D^+\Phi_1^+D^-\Phi_1^- \cr
 &\quad + { \alpha_0 \over 8 }
\del\Phi_2^- D^+\Phi^+_1 D^-\Phi_1^-
 +{ \alpha_0\over 8 } \del \Phi_1^- D^+\Phi_2^+ D^-\Phi_2^-  + {\alpha_0
\over 8 } \del \Phi_2^-
D^+\Phi_2^+ D^- \Phi_2^- \cr
 &\quad  + {  \alpha_0 \over 8 }\del\Phi_1^- D^-\Phi_2^- D^+\Phi_1^+ -{\alpha_0
\over 8 } \del\Phi_1^+
D^+\Phi_2^+D^-\Phi_2^-\cr
&\quad + { 1 \over 16 } D^+\Phi^+_1 D^-\Phi^-_1D^+\Phi^+_2
D^-\Phi_2^-\ .&(2.11b)\cr }
$$
These supercurrents have (quasi) conformal spins 1 and 2 respectively, and
may be expanded in components as
$$
\eqalign{
T(z,\theta^+,\theta^-)&=\ft12 J(z) -\ft12 \theta^+ G^-(z) +\ft12 \theta^-
G^+(z) + \theta^+\theta^- T(z),\cr
{\widetilde W}(z,\theta^+,\theta^-)&={\widetilde V}(z) +\theta^+
{\widetilde U}^-(z) + \theta^- {\widetilde U}^+(z) +\theta^+ \theta^-
{\widetilde W}(z).\cr}\eqno(2.12)
$$
The component fields $(J,G^-,G^+,T)$, and $({\widetilde V},{\widetilde U}^-,
{\widetilde U}^+,{\widetilde W})$ have (quasi) conformal spins
$(1,\ft32,\ft32,2)$ and $(2,\ft52,\ft52,3)$ respectively, as measured by the
energy-momentum tensor $T(z)$.  The tildes on the spin-2 $N=2$
supermultiplet denote the fact that the corresponding component fields are
only quasi-primary, and not primary.  In subsequent sections, we shall find
it convenient to add terms built from the component fields of the $T$
multiplet to the components of the $W$ multiplet, in order to construct
primary currents.  However, for the present purposes of simply exhibiting a
realisation of the super-$W_3$ algebra, this point is inessential, and we
shall postpone further discussion of it for now.  The detailed expressions
for the primary currents are given in the Appendix.

\bigskip
\noindent
{\bf 3. Multi-field Realisations for Super-$W_3$ Strings}
\bigskip

     In the case of the bosonic $W_3$ algebra, the problem of constructing
multi-scalar realisations was considered in [4].  The approach taken in [4]
was to write down an ansatz for the most general possible $W_3$ currents
that could be built from $n$ scalars, and then solve the conditions on the
undetermined coefficients that result from demanding that the currents
generate the algebra.  The final conclusion in that case was that
realisations could be obtained for arbitrary numbers of scalars [4].  A
remarkable feature of all of these realisations is that they can be viewed
as being built from one special scalar, say $\varphi_1$, together with an
energy-momentum tensor ${\cal T}$ for the remaining $(n-1)$ scalars (which
may have background charges).  Thus all but one of the scalars in the
$n$-scalar realisation of $W_3$ enter only {\it via} their stress tensor
[4].  In fact, the only relevant property of the stress tensor ${\cal T}$ is
that it must generate the Virasoro algebra with an appropriate central
charge, which must be correlated with the background charge for the
distinguished scalar $\varphi_1$.  The details of what is ``inside'' ${\cal
T}$ are otherwise inessential.  The multi-scalar realisations can in fact be
built by taking the standard two-scalar Miura realisation for $W_3$, and
replacing the terms involving the second scalar $\varphi_2$ of that
realisation, which appears only {\it via} its stress tensor, by a general
stress tensor ${\cal T}$ with equivalent ``external'' characteristics.

     It is natural, therefore, to look for multi-field realisations for
the $N=2$ super-$W_3$ algebra by using a similar approach.  As we shall
show, the essential feature that enables us to find multi-field realisations
is that it turns out that the superfields $\Phi^+_1$ and $\Phi^-_1$ in
(2.11$a$,$b$) enter the $T$ and ${\widetilde W}$ supercurrents only {\it
via} their super stress tensor.  Thus, in a manner analogous to that for the
bosonic $W_3$ case, one may then replace this super stress tensor by a more
general one with equivalent external characteristics.

     In an obvious notation, we may write the $T$ supercurrent (2.11$a$) as
the sum of terms:
$$
T(z,\theta^+,\theta^-)=T_1 +T_2,\eqno(3.1)
$$
where $T_1$ denotes the part that involves only the $\Phi^\pm_1$ fields, and
$T_2$ denotes the part that involves only the $\Phi^\pm_2$ fields.  Note
that since $\Phi^\pm_1$ and $\Phi^\pm_2$ commute, it follows that $T_1$ and
$T_2$ commute.  Each independently generates the $N=2$ superconformal
algebra, with central charges
$$
\eqalign{
c_1&=3(1+2\alpha_0^2),\cr
c_2&=3(1+4\alpha_0^2),\cr}\eqno(3.2)
$$
respectively.

     We find that the ${\tW}$ supercurrent can be written as:
$$\eqalign{
\tW &= -\ft14\alpha_0^3\del^2\Phi_2^-+\ft14\alpha_0^2(\del\Phi_2^-)^2-\ft14
\alpha_0^2\del\Phi_2^+\del\Phi_2^-+\ft18\alpha_0\del\Phi_2^- D^+\Phi_2^+D^-
\Phi_2^-\cr
&\quad+\ft12\alpha_0^2\del T_1-\ft12\alpha_0\del\Phi_2^-T_1+\ft12\alpha_0\del
\Phi_2^+T_1+\ft14\alpha_0
D^-\Phi_2^-D^+T_1-\ft14D^+\Phi_2^+D^-\Phi_2^-T_1\ ,\cr}\eqno(3.3)
$$
where products of the supercurrent $T_1$ and the superfields are normal
ordered with respect to the superfields in the products and those in the
supercurrent $T_1$. Since $\Phi_1$ and $\Phi_2$ commute, it follows that the
only property of $T_1$ that is relevant is that it generates the $N=2$
superconformal algebra with central charge $c_1$ given by (3.2). Thus we may
generalise to an arbitrary $p$-complex-superfield realisation in which $T_1$
is replaced by a stress tensor ${\cal T}$ with the same central charge, but
built from $p-1$ complex superfields:
$$
{\cal T}=-\ft14D^+\Phi_1^+D^-\Phi_1^-+\ft12\beta_0\del
\Phi_1^+-\ft12\beta_0\del\Phi_1^-
-\ft14\sum_{\mu=3}^p D^+\Phi_\mu^+D^-\Phi_\mu^-\ .\eqno(3.4)
$$
Here we have, without loss of generality, chosen the necessary background
charge to lie in the $\Phi_1$ direction. The background-charge parameter
$\beta_0$ should satisfy
$$
\beta_0^2=\alpha_0^2+\ft12(2-p)\ ,\eqno(3.5)
$$
in order that ${\cal T}$ have the correct central charge $c_1$ given by
(3.2). The total central charge of the $p$-superfield realisation of $N=2$
super-$W_3$ algebra is then
$$
c=6(1+3\alpha_0^2)\ .\eqno(3.6)
$$
The corresponding currents $T(z,\t^+,\t^-)$ and $\tW(z,\t^+,\t^-)$ are given by
$$
\eqalign{
T&=-\ft14 D^+ \Phi^+_2 D^-\Phi^-_2
+{\alpha_0\over2} \partial\Phi^+_2-\alpha_0 \partial\Phi^-_2+{\cal T}\ ,\cr
\cr
\tW &= -\ft14\alpha_0^3\del^2\Phi_2^-+\ft14\alpha_0^2(\del\Phi_2^-)^2-\ft14
\alpha_0^2\del\Phi_2^+\del\Phi_2^-+\ft18\alpha_0\del\Phi_2^- D^+\Phi_2^+D^-
\Phi_2^-\cr
&\quad+\ft12\alpha_0^2\del {\cal T}-\ft12\alpha_0\del\Phi_2^-{\cal T}
+\ft12\alpha_0\del
\Phi_2^+{\cal T}+\ft14\alpha_0
D^-\Phi_2^-D^+{\cal T}-\ft14D^+\Phi_2^+D^-\Phi_2^-{\cal T}\ ,\cr}\eqno(3.7)
$$
with ${\cal T}$ defined by (3.4).  The prescription for replacing $\tW$ by a
primary supercurrent is given in the Appendix.

\np
\noindent
{\bf 4. The Ghost Vacuum and the $L_0$ Intercept}
\bigskip

     We are interested in determining the conditions for anomaly freedom for
the super-$W_3$ algebra.  As we remarked in the introduction, one way to do
this would be to construct the BRST operator, and find the conditions under
which it would be nilpotent.  This would determine the
critical value for the central charge, and also the values for the
intercepts of the bosonic generators in the algebra.
However, since the construction of the BRST operator is likely to be very
complicated, we prefer for now to obtain the anomaly-free conditions by
other means.  For the intercepts, we shall do this by first using a
ghost-counting argument to obtain the value of the $L_0$ intercept for the
Virasoro generators.  Then, in the next section, by contructing physical
states with zero norm (``spurious states''), we shall obtain results for the
intercepts of the remaining bosonic generators.  We shall also give an
argument that determines the value of the critical central charge.

     The calculation of the $L_0$ intercept can in fact be performed quite
generally for a wide class of extended conformal algebras.  In [11], results
were obtained for all the $W_n$ algebras.  In this section, we begin by
examining the $W_n$ case, and then we go on to discuss the generalisation
to the $N=2$ super-$W_n$ algebras.

     The physical states in string theory or $W_n$ string theory have the form
$$
\big|{\rm phys}\big\rangle =\big|p\big\rangle_{\rm mat} \otimes \big|0
\big\rangle_{\rm gh},\eqno(4.1)
$$
where $\big|p\big\rangle_{\rm mat}$ denotes physical states in the
matter sector, and $\big|0\big\rangle_{\rm gh}$ denotes the ghost vacuum.
Physical states must satisfy the condition that $L_m\big|{\rm phys}\big\rangle
=0$ for $m\ge 0$, where $L_m=L_m^{\rm mat}+L_m^{\rm gh}$ is the direct sum
of Virasoro generators in the matter and ghost sectors.  Thus the intercept
$a$ in the matter sector, defined by $(L^{\rm mat}_0-a) \big|p\big\rangle_
{\rm mat}=0$, can be calculated from
$$
L^{\rm gh}_0\big|0\big\rangle_{\rm gh}=-a\big|0\big\rangle_{\rm
gh}.\eqno(4.2)
$$

      The ghost vacuum $\big|0\big\rangle_{\rm gh}$ is
constructed from the $SL(2,C)$-invariant vacuum $\big|0\big\rangle$.  For
the $W_n$ algebra, which has currents of each spin $2\le s\le n$, one has a
ghost field $c^{(s)}(z)$ and an antighost field
$b^{(s)}(z)$ for each current.  These may be expanded in
terms of their Laurent modes $c^{(s)}_m$ and $b^{(s)}_m$,  which satisfy the
anti-commutation relations $\{b^{(s)}_p,c^{(s')}_q\}=\delta^{s
s'}\delta_{p+q,0}$:
$$
\eqalign{
c^{(s)}(z)&=\sum_n c^{(s)}_n z^{-n+s-1},\cr
b^{(s)}(z)&=\sum_n b^{(s)}_n z^{-n-s}.\cr}\eqno(4.3)
$$
The $SL(2,C)$-invariant vacuum is defined by
$$
\eqalign{
b^{(s)}_p\big|0\big\rangle&=0,\quad p\ge -s+1,\cr
c^{(s)}_p\big|0\big\rangle&=0,\quad p\ge s.\cr}\eqno(4.4)
$$
The lowest-energy state, the true ghost vacuum $\big|0\big\rangle_{\rm gh}$,
can be built by acting on the $SL(2,C)$-invariant vacuum with the product of
all the $c^{(s)}_p$ modes with positive $p$ that do not annihilate $\big| 0
\big\rangle$:
$$
\big|0\big\rangle_{\rm gh}\equiv \prod_{s=2}^n \prod_{p=1}^{s-1} c^{(s)}_p
\big|0\big\rangle.\eqno(4.5)
$$
This ghost vacuum has the property, as it must, that
$$
b^{(s)}_p\big|0\big\rangle_{\rm gh}=0\ ,\qquad
c^{(s)}_p\big|0\big\rangle_{\rm gh}=0\ ,
\quad p\ge 1\ ,\eqno(4.6)
$$
whilst $b^{(s)}_p$ and $c^{(s)}_p$ do not annihilate the ghost vacuum if
$p\le -1$.

     Note that we can rewrite (4.5) as
$$
\big|0\big\rangle_{\rm gh}\equiv \prod_{s=2}^n \prod_{p=0}^{s-2}\del^p c^{(s)}
(0)
\big|0\big\rangle,\eqno(4.7)
$$
where the zero argument indicates that we set the argument $z$ of the
normal-ordered product of ghost fields to zero.   This expression can now be
written in a very elegant form, by bosonising the $(b,c)$ ghost systems.  We
do this by introducing a real scalar field $\phi^{(s)}$ for the bosonised
($b^{(s)},c^{(s)}$) ghost system for each bosonic spin $s$ current, and
using the bosonisation rule
$$
b^{(s)}\rightarrow e^{-i\phi^{(s)}},\qquad c^{(s)}\rightarrow
e^{i\phi^{(s)}}.\eqno(4.8)
$$
Since these must have conformal spins $s$ and $(1-s)$ respectively, it follows
that the energy-momentum tensor for $\phi^{(s)}$ takes the form
$$
T_{\phi}^{(s)}=-\ft12(\del\phi^{(s)})^2 +i Q_s\del^2\phi^{(s)},\eqno(4.9)
$$
where the background-charge parameter $Q_s$ is given by
$$
Q_s=s-\ft12.\eqno(4.10)
$$
The scalar fields $\phi^{(s)}$ satisfy the OPEs
$$
\phi^{(s)}(z)\phi^{(s')}(w)\sim -\delta^{ss'}\log(z-w).\eqno(4.11)
$$
The central charge for $T_\phi^{(s)}$ is $c=1-12Q_s^2$.  This is equal to
$-2(6s^2-6s+1)$, which is the correct central charge for $T_{bc}=-s b\del
c+(1-s)\del b c$, the energy-momentum tensor for a spin-$s$ $(b,c)$ system.

     For fields satisfying $\phi(z)\phi(w)\sim -\log(z-w)$, one can show that
$$
e^{a\phi(z)}\ e^{b\phi(w)}\sim (z-w)^{-ab} e^{(a+b)\phi(w)} +
O((z-w)^{-ab+1}).\eqno(4.12)
$$
(Normal ordering is understood; for example, $e^{a\phi(z)}$ is a shorthand
notation for $:e^{a\phi}:(z)$.)  From this, one can prove that
$$
e^{ip\phi}\del^p e^{i\phi}=(-)^p p! e^{i(p+1)\phi}\ .\eqno(4.13)
$$
Applying this iteratively, we find that $\prod_{p=0}^{s-2}\del^p c^{(s)}$ may
be rewritten in bosonised form as $e^{i(s-1)\phi^{(s)}}$, up to an irrelevant
constant factor.  Thus, using (4.10), we may write (4.7) as
$$
\eqalign{
\big|0\big\rangle_{\rm gh}&=\prod_{s=2}^n e^{i(Q_s-\ft12)\phi^{(s)}(0)}
\big|0\big\rangle,\cr
&=e^{i\phi^{(2)}(0)}e^{2i\phi^{(3)}(0)}\cdots e^{i(n-1)
\phi^{(n)}(0)}\big|0\big\rangle\ .\cr}\eqno(4.14)
$$

     Since $[L_0,c^{(s)}_p]=-p c^{(s)}_p$, it follows from (4.5) that the
constant $a$ in (4.2) is given by
$$
\eqalign{
a&=\sum_{s=2}^n\sum_{p=1}^{s-1} p\cr
&=\ft16 n(n^2-1).\cr}\eqno(4.15)
$$
This, then, is the value for the $L_0$ intercept for the $W_n$ algebra [11].
If we consider $n=2$, which is the Virasoro algebra, we recover
the standard result $a=1$; for $n=3$, we obtain the result $a=4$, given in
[2].

     To extend our discussions to supersymmetric algebras, we have to
generalise the arguments to include the commuting $(\beta,\gamma)$ ghost
systems for the fermionic currents. For the $N=1$ super-Virasoro algebra,
this is most easily done by bosonising the $(\beta,\gamma)$ ghosts for the
spin-$\ft32$ current $G(z)$ in
terms of a scalar field $\sigma$ and a pair of anti-commuting fields $\eta$
and $\xi$ with spins 1 and 0 respectively:
$$
\beta\rightarrow\del\xi e^{-\sigma},\qquad  \gamma\rightarrow\eta e^\sigma
\ ,\eqno(4.16)
$$
where the fields satisfy the OPEs
$$
\eta(z)\xi(w)\sim {1\over z-w},\qquad \sigma(z)\sigma(w)\sim
-\log(z-w).\eqno(4.17)
$$
The stress tensor
$T_{\beta\gamma}=-\ft32\beta\del\gamma-\ft12\del\beta\gamma$ becomes
$$
T_{\sigma\eta\xi}=-\ft12(\del\sigma)^2-\del^2\sigma-\eta\del\xi\ .\eqno(4.18)
$$
{}From this we see that $e^{-\sigma}$ has spin $\ft12$, and $e^\sigma$ has spin
$-\ft32$. One can easily verify that the central charges of $T_{\beta\gamma}$
and $T_{\sigma\eta\xi}$ have the same value, namely $c=11$.

     The ghost vacuum for the $N=1$ super-Virasoro algebra is obtained from the
$SL(2,C)$-invariant vacuum $\big|0\big\rangle$ as follows [12]:
$$
\big|0\big\rangle_{\rm
gh}=c^{(2)}(0)e^{-\sigma(0)}\big|0\big\rangle.\eqno(4.19)
$$
This has the required property that $\beta_r$ and $\gamma_r$ annihilate it
for $r \ge \ft12$, but not for $r \le -\ft12$.  Acting with the total ghost
stress tensor $T_{\rm gh}$, and looking at the $O((z-w)^{-2})$ term, one
finds that the ghost vacuum has $L^{\rm gh}_0$ eigenvalue $-1+\ft12=-\ft12$.
 Thus the $L_0$ intercept for the $N=1$ super-Virasoro algebra is $a=\ft12$.
 For the $N=2$ super-Virasoro algebra, we need two ghost pairs
$(\beta,\gamma)$ and $(\bar\beta,\bar\gamma)$ for the two spin-$\ft32$
currents $G^+(z)$ and $G^-(z)$, and the corresponding ghost vacuum
$$
\big|0\big\rangle_{\rm gh}=c^{(2)}(0)e^{-[\sigma(0)+\bar\sigma(0)]}
\big|0\big\rangle\eqno(4.20)
$$
will have $L^{\rm gh}_0$ eigenvalue $-1+\ft12+\ft12=0$.  Thus the $L_0$
intercept for the $N=2$ super-Virasoro algebra is $a=0$.

     For the $N=2$ super-$W_n$ algebra, we must introduce a $(b,c)$ ghost
system for each bosonic current represented in (1.4), and a $(\beta,\gamma)$
ghost system for each fermionic current in (1.4).  The ghosts
$\beta^{(s)}(z)$, $\gamma^{(s)}(z)$ and $\bar\beta^{(s)}(z)$,
$\bar\gamma^{(s)}(z)$ for the complex-conjugate pair of spin-$s$ fermionic
currents can then be bosonised in terms of a scalar $\sigma^{(s)}$ and a
pair of anticommuting fields $\eta^{(s)}$ and $\xi^{(s)}$ with spins 1 and 0
respectively, together with their conjugate set. Thus we have
$$
\beta^{(s)}\rightarrow \del\xi^{(s)}e^{-\sigma^{(s)}}, \qquad
\gamma^{(s)}\rightarrow \eta^{(s)} e^{\sigma^{(s)}}.\eqno(4.21)
$$
The stress tensor $T_\beta\gamma^{(s)}=-s \beta^{(s)} \del \gamma^{(s)} + (1-s)
\del \beta^{(s)} \gamma^{(s)}$ becomes
$$
T_{\sigma\eta\xi}^{(s)}=-\ft12(\del\sigma^{(s)})^2 -Q_s \del^2 \sigma^{(s)}
-\eta^{(s)}\del \xi^{(s)},\eqno(4.22)
$$
where the background charge $Q_s$ is given by
$$
Q_s=s-\ft12.\eqno(4.23)
$$
(Note that this is the same expression, now used for half-integer $s$, as
that obtained in the bosonisation of the $b^{(s)},c^{(s)}$ system (4.10).)
Thus the operators $e^{-\sigma^{(s)}}$ and $e^{\sigma^{(s)}}$ have spins
$s-1$ and $-s$ respectively.  One can check that $T_{\beta\gamma}^{(s)}$ and
$T_{\sigma\eta\xi}^{(s)}$ have the same central charge
$c=12s^2-12s+2=(1+12Q_s^2)-2$. (A similar discussion applies, {\it mutatis
mutandis}, for the ghost pairs $(\bar\beta^{(s)}, \bar\gamma^{(s)})$.)

     We saw in (4.19) that the correct ghost vacuum for the $N=1$
super-Virasoro algebra involves acting on the $SL(2,C)$-invariant vacuum
with the operator $e^{-\sigma(0)}$.  For the $N=2$ super-$W_n$ case, it
turns out that we should act on the $SL(2,C)$-invariant vacuum with a factor
$e^{-Q_s \sigma^{(s)}(0)}$ for each spin-$s$ fermionic current represented
in (1.4), {\it i.e.}, in all, a factor
$$
e^{-\sigma^{(3/2)}(0)}e^{-2\sigma^{(5/2)}(0)}\cdots
e^{-(n-1)\sigma^{(n-1/2)}(0)},\eqno(4.24)
$$
multiplied by a similar factor with scalars $\bar\sigma^{(s)}$, since there
are two real fermionic currents at each spin.
This structure for the fermionic factors in the ghost vacuum is dictated by
the fact that the ghost vacuum $\big|0\big\rangle_{\rm gh}$ should be
annihilated by $\beta^{(s)}_r$ and $\gamma^{(s)}_r$ for all $r\ge \ft12$,
whilst all $\beta^{(s)}_r$ and $\gamma^{(s)}_r$ with $r\le-\ft12$ should not
annihilate it.  This requires that the coefficients of $-\sigma^{(s)}$ in the
exponentials must be precisely equal to the background charges $Q_s$.  The
operator $e^{-Q_s\sigma^{(s)}}$ has conformal spin $\ft12 Q_s^2$, measured
using (4.22).  For each bosonic current of spin $s$ represented in (1.4), we
should act with a factor of $e^{i(Q_s-\ft12)\phi^{(s)}(0)}$, as in (4.14).
This operator has conformal spin $-\ft12Q_s^2+\ft18=-\ft12s(s-1)$.  Thus if
we collect the results for the contributions coming from the ghosts for all
the fields in an $N=2$ supermultiplet, represented by the lozenge
$$
\left\{\matrix{&(s+\ft12)&\cr
               s&&(s+1)\cr
               &(s+\ft12)&\cr}\right\},\eqno(4.25)
$$
we find that the total conformal spin of the operators for this
supermultiplet that act on the $SL(2,C)$-invariant vacuum is given by
$$
\Big(-\ft12 Q_s^2+\ft18\Big) +2\times \Big(\ft12 Q_{s+\ft12}^2\Big)
+\Big( -\ft12 Q_{s+1}^2+\ft18\Big) =0.
\eqno(4.26)
$$
Thus each $N=2$ supermultiplet contributes 0 to the $L^{\rm gh}_0$
eigenvalue, and so the $L_0$ intercept for any of the $N=2$ super-$W_n$
algebras is given by $a=0$.  The corresponding ghost vacuum is given by
$$
\big|0\big\rangle_{\rm gh}= \Big(\prod_{s=2}^n
e^{i(Q_s-\ft12)\phi^{(s)}(0)}
\Big) \Big(\prod_{s=1}^{n-1}e^{i(Q_s-\ft12)\tilde\phi^{(s)}(0)}
\Big) \Big(\prod_{s=\ft32}^{n-\ft12} e^{-Q_s[
\sigma^{(s)}(0)+\bar\sigma^{(s)}(0)]}\Big) \big|0\big\rangle, \eqno(4.27)
$$
where $\phi^{(s)}$ denotes the bosonised ghost fields for the bosonic
currents sitting on the right-hand side of each lozenge in (1.4), and
$\tilde\phi^{(s)}$ denotes the bosonised ghost fields for the bosonic
currents sitting on the left-hand side of each lozenge.  The fields
$\sigma^{(s)}$ and $\bar\sigma^{(s)}$ are the scalars in the bosonisation of
the $(\beta,\gamma)$ and $(\bar\beta,\bar\gamma)$ ghost systems for the
fermionic currents in each lozenge.

\bigskip
\noindent
{\bf 5. The Intercepts for the Super-$W_3$ Algebra}
\bigskip

    In the previous section we showed that the $L_0$ intercepts could
be calculated for the bosonic $W_n$ algebras and the $N=2$ super-$W_n$
algebras, by studying the structure of the ghost vacuum. In principle,
one could use the same method to calculate the intercepts for
higher-spin generators in the algebras. However, because of the nonlinearities
of these algebras, the structures of the higher-spin ghost currents are
complicated, involving products of ghost fields and matter currents.
Therefore we choose to adopt a different method to calculate the higher-spin
intercepts. Our principal goal in this section is to find the intercepts
for the $N=2$ super-$W_3$ algebra.

     The idea of this method is that one may deduce the intercepts, which
are the eigenvalues of the zero modes of the bosonic currents acting on
physical
states, by choosing certain simple examples of physical states, namely null
states. The basic idea has been used, for example in [13], in the case of the
Virasoro and the super-Virasoro algebras. In those cases, the method could
be used at the level of the abstract algebra since the commutation
relations are comparatively simple. In the case of the super-$W_3$ algebra,
however, the complexities of the highly-nonlinear commutation relations [9]
make this approach rather unattractive. Instead, we find that it is easier
to work with a specific realisation of the algebra. In fact we shall use
the Miura realisation that we discussed in section 2.  It should be
emphasised, however, that the use of a specific realisation is merely a
device for obtaining results that could equally well be obtained (in
principle) by using the commutation relations of the abstract algebra.
To illustrate how this method works, we shall first use it to calculate the
$W_0$ intercept for the bosonic $W_3$ algebra.

     The Miura transformation gives a realisation of $W_3$ in terms of two
real scalars [3]:
$$
\eqalignno{
T&=\ft12 (\partial\varphi_1)^2+\ft12(\partial\varphi_2)^2 +
\big( a_1\partial^2\varphi_1 +a_2\partial^2\varphi_2
 \big)\ ,&(5.1a)\cr\cr
W&=\ft13(\partial\varphi_1)^3
-\partial\varphi_1 (\partial\varphi_2)^2 +\big(a_1
\partial\varphi_1  \partial^2\varphi_1 -2 a_2 \partial\varphi_1
\partial^2\varphi_2 - a_1 \partial\varphi_2 \partial^2 \varphi_2 \big)\cr
&\ \ +\big( \ft13 a_1^2\partial^3\varphi_1
-a_1 a_2 \partial^3\varphi_2 \big)
\ ,&(5.1b)\cr}
$$
where $a_2={1\over{\sqrt 3}}a_1$, and the central charge is given by $c=2+16
a_1^2$. One example of a null physical state is $\big|p\big\rangle\equiv
P(0) \big|0\big\rangle_{\rm mat}$, where $\big|0\big\rangle_{\rm mat}$
stands for the SL(2,C)-invariant matter vacuum, and
$$
P(z)\equiv (L_{-1}+gW_{-1})e^{\beta\cdot\varphi(z)}\ .\eqno(5.2)
$$
Here $g$ is a constant parameter and $\beta\cdot\varphi=\beta_1\varphi_1
+\beta_2\varphi_2$.
Note that $L_{-1}e^{\beta\cdot\varphi}$ can be obtained as the coefficient
of the $1\over{(z-w)}$ term in the OPE of $T(z)e^{\beta\cdot\varphi(w)}$.
Similarly, $W_{-1}e^{\beta\cdot\varphi}$ can be obtained as the coefficient
of the $1\over{(z-w)^2}$ term in the OPE of $W(z)e^{\beta\cdot\varphi(w)}$.

      Defining the eigenvalues of $L_0$ and $W_0$ acting on
$e^{\beta\cdot\varphi(0)}$ by
$$
\eqalign{
L_0 e^{\beta\cdot\varphi(0)}&=\Delta e^{\beta\cdot\varphi(0)}\ ,\cr
W_0 e^{\beta\cdot\varphi(0)}&=\omega e^{\beta\cdot\varphi(0)}\ ,\cr}\eqno(5.3)
$$
we find that
$$\eqalign{
\Delta&=-a_1\beta_1-a_2\beta_2+\ft12\beta_1^2+\ft12\beta_2^2\ ,\cr
\omega&=\ft13(\beta_1-a_1)(\beta_1^2-3\beta_2^2+6a_2\beta_2-2a_1\beta_1)\ .\cr}
\eqno(5.4)
$$
Physical states must be annihilated by $L_m$ and $W_m$ with $m\ge1$.  On the
state defined by (5.2), the only non-trivial conditions are $L_1P(0)=0$ and
$W_1P(0)=0$.  These give
$$\eqalign{
2\Delta+3g\omega&=0\ ,\cr
3\omega+2g(a_1^2+2\Delta)\Delta&=0\ .\cr}
\eqno(5.5)
$$
Defining the eigenvalues $L_0$ and $W_0$ acting on $P(0)$ to be $a$ and
$\lambda$, given by
$$
\eqalign{
L_0P(0)&=aP(0)\ ,\cr
W_0P(0)&=\lambda P(0)\ ,\cr}\eqno(5.6)
$$
we find that they are related to $\Delta$ and $\omega$ by
$$
\eqalign{
a&=1+\Delta\ ,\cr
\lambda&=\omega+{2\over g}\ .\cr}\eqno(5.7)
$$
But, as we showed in (4.15), the $L_0$ intercept is $a=4$ for the $W_3$
algebra, which implies from (5.7) that $\Delta=3$ and hence from (5.5) that
$\omega=-{2/g}$. It then follows from (5.7) that $\lambda=0$; in other
words, the intercept for $W_0$ is zero. Note that this result is obtained
without fixing any specific value for the background charge. Thus
the intercepts are determined by this method without requiring that the
anomaly-free condition for the central charge $c$ be imposed.

     This method can also be applied to the $N=2$ super-$W_3$ algebra. In
this case, we use the super-Miura realisation given in section 2. The
explicit expressions for the currents in terms of components are given in
the Appendix. The component fields in this realisation comprise the bosons
$\phi_1, {\bar\phi}_1, \phi_2, {\bar\phi}_2$ and fermions $\psi_1,
{\bar\psi}_1, \psi_2, {\bar\psi}_2$. The OPE's for these fields are
given in (A.4). The simplest null physical states for this realisation can be
constructed as $\big|p\big\rangle\equiv P(0) \big|0\big\rangle_{\rm mat}$,
where
\cramp
$$
\eqalign{
P(z)&\equiv -{1\over{\sqrt 2}}\Big( G_{-1/2}^++gU_{-1/2}^+\Big)
e^{\beta\cdot{\bar\phi}+{\bar\beta}\cdot\phi}\cr\cr
&=\Big[{\bar\beta}_1\big(1-{g\over4}(\ft23\alpha_0^2+\alpha_0\beta_2)\big)
\psi_1+{\bar\beta}_2\big(1-{g\over
4}(-\ft13\alpha_0^2+\alpha_0\beta_1-\alpha_0{\bar\beta}_1-\alpha_0{\bar\beta}_2)
\big)\psi_2\Big]
e^{\beta\cdot{\bar\phi}+{\bar\beta}\cdot\phi}\ .\cr}
\eqno(5.8)
$$
\uncramp

     It is convenient to introduce the quantities $j,\ \Delta,\ \nu$ and
$\omega$, defined as the eigenvalues of $J_0$, $L_0$, $V_0$ and $W_0$
acting on the
operator $e^{\beta\cdot{\bar\phi(0)}+{\bar\beta}\cdot\phi(0)}$:
$$
\eqalign{
J_0 e^{\beta\cdot{\bar\phi(0)}+{\bar\beta}\cdot\phi(0)}&=j
e^{\beta\cdot{\bar\phi(0)}+{\bar\beta}\cdot\phi(0)}\ ,\cr
L_0 e^{\beta\cdot{\bar\phi(0)}+{\bar\beta}\cdot\phi(0)}&=\Delta
e^{\beta\cdot{\bar\phi(0)}+{\bar\beta}\cdot\phi(0)}\ ,\cr
V_0 e^{\beta\cdot{\bar\phi(0)}+{\bar\beta}\cdot\phi(0)}&=\nu
e^{\beta\cdot{\bar\phi(0)}+{\bar\beta}\cdot\phi(0)}\ ,\cr
W_0 e^{\beta\cdot{\bar\phi(0)}+{\bar\beta}\cdot\phi(0)}&=\omega
e^{\beta\cdot{\bar\phi(0)}+{\bar\beta}\cdot\phi(0)}\ .\cr}
\eqno(5.9)
$$
Using the realisation in Appendix, we find after some algebra that these
eigenvalues can be written in terms of the momentum components
$\beta_1,\bar\beta_1,\beta_2,\bar\beta_2$ as
$$\eqalign{
j&=-\alpha_0\beta_1-\alpha_0\beta_2+\alpha_0{\bar\beta}_1+2\alpha_0{\bar\beta}_2
\ ,\cr
\Delta&=-\ft12\alpha_0\beta_1-\ft12\alpha_0\beta_2-\ft12\alpha_0{\bar\beta}_1
-\alpha_0{\bar\beta}_2-\beta_1{\bar\beta}_1-\beta_2{\bar\beta}_2\ ,\cr
\nu&=\ft1{12}\a^2\Big(2\a\b_1-\a\b_2-\a\bb_1+\a
\bb_2\cr
&\qquad\quad -3\b_1\bb_2-2\b_2\bb_2+3\b_1\b_2+\b_1\bb_1
-3\b_2\bb_1+3\bb_1\bb_2+3\bb_2^2\Big)\ ,\cr
\omega&=\omega'-{{3+8\a^2}\over{4(4+9\a^2)}}j\Delta\ ,\cr}
\eqno(5.10)
$$
where
$$\eqalign{
\omega'=-\ft16\a\Big(&-2\a^2\b_1+\a^2\b_2-\a^2\bb_1+\a^2\bb_2\cr
& -3\a\b_1\b_2 -3\a\b_1\bb_1-3\a\b_1\bb_2+3\a\b_2\bb_2+3\a\bb_1\bb_2
+3\a\bb_2^2\cr
& -3\b_1\b_2\bb_1-3\b_1\b_2\bb_2+3\b_2\bb_1\bb_2+3\b_2\bb_2^2\Big)\ .
\cr}
\eqno(5.11)
$$

     The physical operator defined by (5.8) must satisfy the physical-state
conditions, namely $G_r^-P(0)=0$ and $U_r^-P(0)=0$ for $r\ge\ft12$. The only
non-trivial conditions in this case will therefore be $G_{1/2}^- P(0)=0$ and
$U_{1/2}^- P(0)=0$.  These give
$$\eqalignno{
0&=2(\Delta-\ft12 j)+ g(\omega'-2\nu)\ , &(5.12a)\cr
0&=-3\a^2g(\Delta-\ft12 j)^2+2\a^4g(\Delta-\ft12 j)+6
(12-\a^2g)(\omega'-2\nu)\cr
&\quad -36 g\nu(\Delta-\ft12 j)+18 gj(\omega'-2\nu)\ .&(5.12b)\cr}
$$
{}From these equations, it follows that there is a solution which is
independent of $\a$ and $g$, with
$$
\Delta-\ft12 j=0\ ,\qquad {\rm and}\qquad \omega'-2\nu=0\ .\eqno(5.13)
$$
(There is an alternative solution of these equations, but as we shall
see later, (5.13) is the relevant one.)

     The physical operator $P(0)$ must be an eigenstate of the zero modes of
all the bosonic operators.  Thus we may define $\rho,\ a,\ \mu,\ \lambda$ as
the eigenvalues for $J_0,\ L_0 ,\ V_0,\ W_0$ on the physical operator
$P(0)$; {\it i.e.}
$$
\eqalign{
J_0P(0)&=\rho P(0)\ ,\cr
L_0P(0)&=a P(0)\ ,\cr
V_0P(0)&=\mu P(0)\ ,\cr
W_0P(0)&=\lambda P(0)\ .\cr}\eqno(5.14)
$$
The physical state defined by (5.8) is automatically an eigenstate of $J_0$
and $L_0$ for all values of $g$, and we find
$$
\eqalign{
\rho&=j+1\ ,\cr
a&=\Delta+\ft12\ .\cr}\eqno(5.15)
$$
For $P(0)$ to be an eigenstate of $V_0$, we find that either the momenta
must satisfy the condition
$$
\a-\beta_1+\beta_2+\bar\beta_1+\bar\beta_2=0,\eqno(5.16a)
$$
or the constant $g$ in (5.8) must satisfy one or other of the conditions
$$
-12+2\a^2 g+3\a g\b_2 -3\a g\bb_2=0\eqno(5.16b)
$$
or
$$
12+\a^2 g-3\a g\b_1+3\a g\bb_1+3\a g\bb_2=0\ .\eqno(5.16c)
$$
For the second possibility, (5.16$b$), the corresponding eigenvalue is given
by
$$
\mu =\nu+\ft14 j-\ft1{12}\a^2+{1\over g}\ .\eqno(5.17)
$$
(We shall discuss the possibilities (5.16$a$) and (5.16$c$) later.)

For $P(0)$ to be an eigenstate of $W_0$, we find that either the condition
(5.16$a$) must again be satisfied, or that $g$ must satisfy
$$
18g^2\omega'+3g(12+\a^2g)\Delta+\ft32 g(12-\a^2g)j-2(12+\a^2 g)(-6+\a^2g)
=0\ .\eqno(5.18)
$$
For the case (5.18), the eigenvalue $\lambda$ can then be written in the form
$$
\lambda=\omega'+{2\over g}+{{5(1+2\a^2)}\over{4(4+9\a^2)}} (\Delta+\ft12 j)
-{{(3+8\a^2)}\over{4(4+9\a^2)}}j\Delta -{{(9+40\a^2+36\a^4)}\over
{24(4+9\a^2)}}\ .
\eqno(5.19)
$$

     At this stage in the argument, we recall from section 4 that the $L_0$
intercept $a$ for the $N=2$ super-$W_3$ algebra is 0. Hence from (5.15) we
see that $\Delta=-\ft12$, and then from (5.13) we find $j=-1$. Note that this
implies, from (5.15), that $\rho=0$. In other words, the intercept for $J_0$
is zero. In fact, there is an elementary argument for why the $J_0$
intercept has to be zero. The super-$W_3$ algebra has a charge-conjugation
symmetry [9]:
$$\eqalign{
&J\lra -J,\quad G^+\lra G^- ,\quad T\lra T\ ;\cr\cr
&V\lra -V,\quad U^+\lra U^- ,\quad W\lra W\ .\cr}\eqno(5.20)
$$
This symmetry would be broken if $J_0$ or $V_0$ had a non-vanishing intercept.
Thus we must have
$$
\rho=0\ ,\qquad \mu=0\ .\eqno(5.21)
$$
(This independent argument gives a justification for why (5.13) is the
preferred solution of (5.12$a,b$).)  This argument extends straightforwardly
to show that the intercepts for all the bosonic currents sitting on the
left-hand sides of the lozenges in (1.4) must be zero for any of the $N=2$
super-$W_n$ algebras.

     There is also another way to see why the $J_0$ intercept is zero.
Although the ghost-vacuum argument that we used in section 4 to determine
the $L_0$ intercept does not easily extend to higher-spin currents, because
of the nonlinearities of the algebra, it can be used quite straightforwardly
in the case of the spin-1 current, since the spin-1 ghost current $J^{\rm
gh}$ cannot receive any matter-dependent nonlinear modifications.  The
generator $J^{\rm gh}_0$ counts the difference between the numbers of ghosts
for fermions and conjugate-fermions in the ghost vacuum, and so, since these
are equal, we have $J^{\rm gh}_0=0$ and hence the $J_0$ intercept vanishes.

     If we now take the expression (5.19) for the eigenvalue $\lambda$ of
$W_0$, and use (5.17), together with the results that $\Delta=-\ft12$ and
$j=-1$, we find that the $W_0$ and $V_0$ eigenvalues of $P(0)$ are related by
$$
\lambda-2\mu=0\ .\eqno(5.22)
$$
Combining this with $\mu=0$ in (5.21), coming from the charge-conjugation
symmetry of the algebra, we finally arrive at the conclusion that $\lambda=0$.
Thus we have shown that the intercepts for all four bosonic currents in the
$N=2$ super-$W_3$ algebra are zero; {\it i.e.}\ acting on physical states, we
have
$$
J_0=0,\quad L_0=0,\quad V_0=0, \quad W_0=0.\eqno(5.23)
$$
Note that this result was arrived at by considering special physical
null states of the form (5.8), with the constant $g$ satisfying the condition
(5.16$b$).  Of course the conclusions that one arrives at by looking at any
other physical state must be consistent with this.  In particular, one can
check that indeed for the alternative choices of conditions (5.16$a$) or
(5.16$c$), one arrives at the same conclusions.

\bigskip
\noindent{\bf 6. Conclusions and Discussions}
\bigskip

     In this paper we have shown how the two-complex-superfield realisation
for the $N=2$ super-$W_3$ algebra, obtained from the super-Miura
transformation, may be generalised to give realisations in terms of an
arbitrary number of complex superfields.  Such realisations will be of
importance for constructing super-$W_3$ strings.  In order to build
super-$W_3$ string theory, one also needs to know the conditions for anomaly
freedom for the algebra; in other words one needs to know the intercepts for
the bosonic generators, and the critical central charge, which would be
necessary in order to have a nilpotent BRST operator.  We have studied the
structure of the ghost vacuum for the $N=2$ super-$W_n$ algebras, and have
shown how one may deduce from this that the $L_0$ intercept is zero for all
$n$.  Then, by considering certain especially-simple physical states that
are null, we have shown by explicit computation that all the other bosonic
currents in the $N=2$ super-$W_3$ algebra have zero intercept too.  It
would be interesting to see whether this result generalises to all the
$N=2$ super-$W_n$ algebras.

     The critical value of the central charge is easily calculated for all
the $N=2$ super-$W_n$ algebras.  We know that the ghosts for a current of
spin $s$ give a contribution of
$$
c_{\rm gh}(s)=(-)^{2s+1}(12s^2-12s+2)\eqno(6.1)
$$
to the total ghost central charge.  Therefore the total contribution from
the ghosts for the fields in the lozenge (4.25) is given by
$$
\eqalign{&-\Big(12s^2-12s+2\Big) +2\Big(12(s+\ft12)^2-12(s+\ft12) +2\Big)
-\Big(12(s+1)^2-12(s+1)+2\Big)\cr
&=-6.\cr}\eqno(6.2)
$$
Thus each lozenge, independently of the spins of its currents, gives a
contribution of $-6$ to the total ghost central charge.  The condition for
anomaly freedom of the $N=2$ super-$W_n$ algebra is therefore that the
central charge $c$ for the matter fields should satisfy
$$
c=6(n-1).\eqno(6.3)
$$
For the $N=2$ super-$W_3$ algebra, we obtain $c=12$, which was already given in
[9].  From (3.6), it follows that we must take $\a^2=\ft13$ for any of the
realisations discussed in section 3.

     These results provide the essential ingredients for the construction
of the $N=2$ super-$W_3$ string. Work on this is in progress [14].

\bigskip\bigskip
     \centerline{\bf ACKNOWLEDGMENTS}
\bigskip

     We are very grateful to Larry Romans for extensive discussions, and
for writing a Mathematica program for calculating the expansion of the
super-Miura transformation (2.3) for the super-$W_n$ algebra.  We have also
made extensive use of a Mathematica package for calculating operator-product
expansions, written by K. Thielemans [15].

\vfill\eject
\centerline{\bf APPENDIX}
\bigskip\bigskip\bigskip

    In section 2, we have given general expressions for the two $N=2$
supercurrents of the super-$W_3$ algebra, namely $T(z,\t^+,\t^-)$ and
$\tW(z,\t^+\t^-)$, in terms of the $N=2$ superfields $\Phi_1^{\pm}$ and
$\Phi_2^{\pm}$. The supercurrent $\tW(z,\t^+\t^-)$ is not primary. Here we
give explicit expressions, in terms of components, for the all the
primary bosonic and fermionic currents.  The first lozenge in (1.3),
corresponding to the the $N=2$ supercurrent $T(z,\t^+,\t^-)$, has component
currents $(J,G^+,G^-,T)$ that are already primary (except that there is, of
course, a central term in the Virasoro algebra).  The component
currents, defined by (2.12), are given by:
$$
\eqalign{
J&=-\psi_1{\bar\psi}_1-\psi_2{\bar\psi}_2+\a (\del\phi_1-\del{\bar\phi}_1
+\del\phi_2-2\del{\bar\phi}_2)\ ,\cr
G^+&=\sqrt 2 (\del{\bar\phi}_1\psi_1+\del{\bar\phi}_2\psi_2+\a\del\psi_1
+\a\del\psi_2)\ ,\cr
G^-&=\sqrt 2 (\del\phi_1{\bar\psi}_1+\del\phi_2{\bar\psi}_2
+\a\del{\bar\psi}_1
+2\a\del{\bar\psi}_2)\ ,\cr
T&=\ft12\psi_1\del{\bar\psi}_1-\ft12\del\psi_1{\bar\psi}_1-\del\phi_1
\del{\bar\phi}_1+\ft12\psi_2\del{\bar\psi}_2-\ft12\del\psi_2{\bar\psi}_2
-\del\phi_2
\del{\bar\phi}_2\cr
&\quad
-\ft12\a\del^2\phi_1
-\ft12\a\del^2{\bar\phi}_1
-\ft12\a\del^2\phi_2
-\a\del^2{\bar\phi}_2\ .\cr}\eqno(A.1)
$$

     For the second lozenge in (1.3), corresponding to the $N=2$
supercurrent $\tW(z,\t^+,\t^-)$, we must add terms involving derivatives and
products of currents from the $T(z,\t^+,\t^-)$ supercurrent in order to
achieve primary currents.  There is some degree of arbitrariness in how to
do this.  Here, we present the specific choices that we have found to be
most useful, and which we have used for our calculations.  Our primary
currents are
$$
\eqalign{
V&={\widetilde V} -\ft18\a^2 \del J -\ft1{12}\a^2 T,\cr
U^+&={\widetilde U}^+ -\ft1{12} \a^2 \del G^+,\cr
U^-&={\widetilde U}^- +\ft16 \a^2\del G^-,\cr
W&={\widetilde W} -\ft1{24}\a^2 \del^2 J -\ft14 \a^2 \del T -{(3+8\a^2)\over
4(4+9\a^2)} J T.\cr}\eqno(A.2)
$$
Note that the product $JT$ is normal ordered with respect to the modes of
$J$ and $T$.

     The non-primary component currents ${\widetilde V}$, ${\widetilde U}^+$,
${\widetilde U}^-$ and ${\widetilde W}$, defined by (2.12), take the
component form:
$$
\eqalignno{
{\widetilde V}&=\ft14\a^3\Big(\del^2\bar\phi_2-\del^2 \phi_2\Big) \cr
&\quad+\a^2\Big( \ft14 \del J+\ft14 \del(\psi_2\bar\psi_2)
-\ft14\del\bar \phi_2
\del\bar\phi_2-\ft14 \del\phi_2 \del\phi_2 +\ft12 \del\phi_2 \del\bar\phi_2
+\ft14 \del\psi_2 \bar\psi_2\Big)\cr
&\quad +\a\Big( \ft14 \del\phi_2 J-\ft14\del\bar\phi_2 J+\ft12 \del\phi_2
\psi_2 \bar\psi_2 -\ft14 \del\bar\phi_2 \psi_2 \bar\psi_2 +\ft18 \sqrt2
\bar\psi_2 G^+\Big) -\ft14 \psi_2 \bar\psi_2 J,\cr
\cr
{\widetilde U}^+&=-\ft14\sqrt2\a^3 \del^2\psi_2 +\sqrt2 \a^2\Big( \ft18\sqrt2
\del G^+ -\ft14 \del(\del\bar\phi_2\psi_2)-\ft12 \del\psi_2 \del\phi_2+
\ft12 \del\bar\phi_2 \del\psi_2-\ft14 \del\psi_2\del\bar\phi_2\Big)\cr
&\quad +\sqrt2\a\Big(\ft18\sqrt2 \del\phi_2
G^+ +\ft14\del\psi_2 J +\ft12\del\psi_2 \psi_2 \bar\psi_2 -\ft12 \del\phi_2
\psi_2 \del\bar\phi_2+\ft14 (\del\bar\phi_2)^2\psi_2\Big)\cr
&\quad -\ft14\psi_2\bar\psi_2 G^+ +\ft14\sqrt2\psi_2\del\bar\phi_2 J,\cr
\cr
{\widetilde U}^-&=\ft14\sqrt2 \a^3\del^2\bar\psi_2 +\sqrt2\a^2\Big(
-\ft18\sqrt2\del G^-
+\ft14\del(\del\phi_2\bar\psi_2)
-\ft12 \del\bar\psi_2 \del\bar\phi_2+\ft12 \del\phi_2\del\bar\psi_2
+\ft14\del^2 \phi_2 \bar\psi_2\Big)\cr
&\quad +\sqrt2 \a\Big(\ft18\sqrt2 \del\bar\phi_2 G^- -\ft18\sqrt2
\del\phi_2 G^- -\ft14\del\bar\psi_2 J +\ft12 (\del\phi_2)^2 \bar\psi_2-\ft14
\del\bar\psi_2 \psi_2 \bar\psi_2\cr
&\quad -\ft14 \del\bar\phi_2 \del\phi_2 \bar\psi_2
+\ft14 \bar\psi_2(T-\ft12\del J) \Big)
+ \ft14 \psi_2\bar\psi_2 G^--\ft14\sqrt2 \del\phi_2 \bar\psi_2 J,\cr
\cr
{\widetilde W}&=\ft14\a^3\Big(\del^3\phi_2+\del^3\bar\phi_2\Big)+
\ft12\a^2\Big(\del T +\del^2\psi_2\bar\psi_2-\ft12 \psi_2\del^2\bar\psi_2
+\ft32 \del\psi_2\del\bar\psi_2\cr
&\quad +\del^2\phi_2 \del\bar\phi_2 +2\del\phi_2
\del^2\bar\phi_2-\del^2\bar\phi_2\del\bar\phi_2
+\del\phi_2\del^2\phi_2\Big)\cr
&\quad+\ft14\a\Big(\ft12\sqrt2 \del\bar\psi_2 G^+-\sqrt2 \del\psi_2 G^-
+\ft12\sqrt2
\bar\psi_2 \del G^+ -\del^2\bar\phi_2 J -2\del\bar\phi_2 T +2 \del\bar
\phi_2(T-\ft12\del J)\cr
&\qquad - \del^2\phi_2 J +2 \del\phi_2 T
+6\del\phi_2\del\psi_2\bar\psi_2 -2\del\phi_2 \psi_2\del\bar\psi_2
-2\del^2\phi_2 \psi_2\bar\psi_2+3\del\bar\phi_2\psi_2\del\bar\psi_2\cr
&\qquad
-\del\bar\phi_2\del\psi_2\bar\psi_2-\del^2\bar\phi_2\psi_2\bar\psi_2
+4(\del\phi_2)^2\del\bar\phi_2-2(\del\bar\phi_2)^2\del\phi_2\Big)\cr
&\quad +\ft14\sqrt2 \del\phi_2\bar\psi_2 G^+ -\ft14\sqrt2 \del\bar\phi_2\psi_2
G^- -\ft14(\del\psi_2\bar\psi_2-\psi_2\del\bar\psi_2+2\del\phi_2\del\bar\phi_2)
J -\ft12\psi_2\bar\psi_2 T.&(A.3)\cr}
$$
Note that terms involving a product of a current and one or more fields are
normal ordered with respect to the fields in the product and those contained
in the currents.

    The OPEs of the fields are
$$\eqalign{
\phi_i(z)\bar\phi_j(w)&\sim-\delta_{ij}\log (z-w)\ ,\cr
\psi_i(z)\bar\psi_j(w)&\sim-{\delta_{ij}\over{(z-w)}},\qquad i,j=1,2\ .\cr}
\eqno(A.4)
$$

     For some purposes it is convenient to work with a choice of primary
fields which can be written into $N=2$ superfield language. The supercurrent
$W'$ given by
$$
W'(z,\t^+,\t^-)={\tW}+b_1
:TT: +b_2\del T
+b_3 \big( D^+D^--D^-D^+\big)
T
\eqno(A.5)
$$
is primary [10], where
$$\eqalign{
b_1&=-{{(3+8\a^2)}\over{2(5+18\a^2)}}\cr
b_2&=-\ft14\a^2\cr
b_3&={{(1+3\a^2)(1-2\a^2)}\over {8(5+18\a^2)}}\ .\cr}\eqno(A.6)
$$

In component language, this becomes
$$\eqalign{
V'(z)&={\tV}+\ft14 b_1 J^2+\ft12 b_2\del J+2b_3 T\ ,\cr
{U'}^+(z)&=\tU^++\ft14 b_1\big(JG^++G^+J\big)+\big(\ft12 b_2-b_3\big)\del G^+
\ ,\cr
{U'}^-(z)&=\tU^--\ft14 b_1\big(JG^-+G^-J\big)-\big(\ft12 b_2+b_3\big)\del G^-
\ ,\cr
W'(z)&=\tW +b_1\big(\ft12 TJ+\ft12 JT+\ft14G^-G^+-\ft14G^+G^-\big)
+b_2\del T+b_3\del^2J\ ,\cr}
\eqno(A.7)
$$
where the products of currents are normal ordered with respect to the modes
of the currents themselves.  This choice of primary currents differs from
the one given in (A.2), which we found convenient to use for the
calculations in section 5.  The choice (A.7) is the unique one that can be
written in $N=2$ superfield language, as in (A.5).

\vfill\eject
\singlespace
\centerline{\bf REFERENCES}
\frenchspacing
\bigskip

\item{[1]}A.B.\ Zamolodchikov, {\sl Teo.\ Mat.\ Fiz.}\ {\bf 65} (1985) 347.

\item{[2]}J.\ Thierry-Mieg, {\sl Phys.\ Lett.}\  {\bf 197B} (1987) 368.

\item{[3]}V.A.\ Fateev and A.\ Zamolodchikov, {\sl Nucl.\  Phys.}\  {\bf
B280} (1987) 644;\nl
V.A.\ Fateev and S.\ Lukyanov,  {\sl Int.\ J.\ Mod.\  Phys.}\ {\bf
A3} (1988) 507.

\item{[4]}L.J.\  Romans, {\sl Nucl.\  Phys.}\ {\bf B352} (1991) 829.

\item{[5]}C.N.\ Pope, L.J.\ Romans and K.S.\ Stelle, {\sl Phys.\
Lett.}\ {\bf 268B} (1991) 167.

\item{[6]}C.N.\ Pope, L.J.\ Romans and K.S.\ Stelle, ``On $W_3$  strings,''
preprint CERN-TH.6171/91, {\sl Phys.\  Lett.}\ {\bf B} (in press).

\item{[7]}C.N.\ Pope, L.J.\ Romans E.\ Sezgin and K.S.\ Stelle, ``The $W_3$
String Spectrum,'' preprint CTP TAMU-68/91.

\item{[8]}H.\ Lu, C.N.\ Pope, L.J.\ Romans, X.\ Shen and X.J.\ Wang, {\sl
Phys.\ Lett.}\ {\bf 264B} (1991) 91.

\item{[9]}L.J.\ Romans, ``The $N=2$ super-$W_3$ algebra,'' preprint
USC-91/HEP06.

\item{[10]}D.\ Nemeschansky and S.\ Yankielowicz, ``$N=2$ $W$-algebras,
Kazama-Suzuki models and Drinfeld-Sokolov reduction,'' preprint USC-91/005.

\item{[11]}S.R.\ Das, A.\ Dhar and S.K.\ Rama, ``Physical properties of $W$
gravities and $W$ strings,'' preprint, TIFR/TH/91-11;\nl ``Physical states
and scaling properties of $W$ gravities and $W$ strings,''\nl
preprint, TIFR/TH/91-20.

\item{[12]}M.E.\ Peskin, ``Introduction to String and Superstring Theory II,''
Lectures presented at the 1986 Theoretical Advanced Study Institute, in
Partcle Physics, UC Santa Cruz.

\item{[13]}M.B.\ Green, J.H.\ Schwarz and E.\ Witten, ``Superstring Theory,''
(CUP 1987).

\item{[14]}H.\ Lu, C.N.\ Pope, X.J.\ Wang and K.W.\ Xu, Work in progress.

\item{[15]}K. Thielemans, ``A Mathematica Package for Computing Operator
Product Expansions,'' preprint, Leuven, May 1991.

\bye